# Understanding differences of the OA uptake within the German university landscape (2010-2020) – Part 1: journal-based OA

Niels Taubert[1] (ORCID iD: 0000-0002-2357-2648), Anne Hobert[2] (0000-0003-2429-2995), Najko Jahn[2] (0000-0001-5105-1463), Andre Bruns[1] (0000-0002-2976-0826), Elham Iravani[1] (0000-0003-1961-2130)

[1] Institute for Interdisciplinary Studies of Science (I2SoS), Bielefeld University, Germany
[2] Göttingen State and University Library, University of Göttingen, Germany

Correspondence: niels.taubert@uni-bielefeld.de

## Abstract

This study investigates the determinants for the uptake of Full and Hybrid Open Access (OA) in the university landscape of Germany. It adapts the governance equaliser as a heuristic for this purpose and distinguishes between three factors: The disciplinary profile (academic self-governance), infrastructures and services of universities that aim to support OA (managerial self-governance) and large transformative agreements (part of state regulation). The uptake of OA, the influence of the disciplinary profile of universities and the influence of transformative agreements is measured by combining several data sources (incl. Web of Science, Unpaywall, an authority file of standardised German affiliation information, the ISSN-Gold-OA 4.0 list, and lists of publications covered by transformative agreements). For *managerial self-governance,* a structured data collection was created by harvesting different sources of information and by manual online search. To determine the explanatory power of the different factors, a series of regression analyses was performed for different periods and for both Full as well as Hybrid OA. As a result of the regression analyses, the most determining factor for the explanation of differences in the uptake of both OA-types turned out to be academic self-governance. For the year 2020, Hybrid OA transformative agreements have become a second relevant factor. However, all variables that reflect local infrastructural support and services for OA (managerial self-governance) turned out to be non-significant. To deepen the understanding of the adoption of OA on the level of institutions, the outcomes of the regression analyses are contextualised by an interview study conducted with 20 OA officers of German universities.

## Declarations

## Funding:
This work was supported by the German Federal Ministry of Education and Research within the funding stream "Quantitative research on the science sector", project OAUNI (grant numbers 01PU17023A and 01PU17023B).

# 1. Introduction

The uptake of Open Access (OA) on the level of institutions has increased in relevance for a number of reasons. Recent science policies and OA strategies have announced that certain percentages of the publication output should be OA at a defined point in time, and universities and other research institutes are committed to such targets. Moreover, funding programmes have been developed that aim to build up infrastructures for the support of OA at remarkable scale, and such programmes are subject to evaluation. One evaluation criterion is, of course, the development of OA in the publication output of participating institutions.[1] Finally, a number of organisations have implemented monitoring and research information systems that aim to measure OA shares on the level of institutions and are supposed to provide information for the further OA transition.[2] On a larger scale, a number of research studies has addressed the question about the dynamics of the uptake of OA. With some exceptions, there is evidence that OA is growing (Archambault et al., 2014; Laakso & Björk, 2012; Piwowar et al., 2018; Pölönen et al., 2020; Severin et al., 2020) with striking differences across institutions (Wohlgemuth et al., 2017; Bosman & Kramer, 2018; Huang et al., 2020) and countries (Robinson-Garcia et al., 2020). In addition, the distinction of different OA types in these studies shows that the uptake of OA is a multidimensional phenomenon (Piwowar et al., 2018; Hobert et al., 2021).

To date, the causes for differences in the uptake of OA are primarily studied on the level of individuals (Bosellii & Galindo-Ruedai, 2016; Tenopir et al., 2017; Rowley et al., 2017; Greussinger et al., 2020), disciplines[3] and countries (Momeni et al., 2022), but rarely on the level of institutions (Greussinger et al., 2020). For institutions, it seems plausible at first glance that different types of factors may play a role. First, there is some evidence that research institutions do not show an equal level of activities when it comes to the support of OA. This does not only hold for services but also for the availability and maintenance of OA infrastructures. Second, the relevant environment, like, for example, guidelines or prescriptions of research policy or guidelines of relevant funding organisations, may be more or less OA-friendly and may result in differences to what extent the publication output of a research organisation is OA. Finally, a number of studies report large differences regarding

---

the OA share in different disciplines and fields.[4] Regarding institutions, the hypothesis seems plausible that differences of the uptake of OA may simply be a reflection of the OA affinity of the disciplinary profile of a research institution. To put it in different words, the question is unanswered whether it is the composition of scientific disciplines, the organisational environment or the inner-organisational support of OA publishing that explains the differences in the uptake of OA.

To the best of our knowledge, our study is the first one that addresses the question of the determinants for the uptake of OA on the level of institutions with a focus on the German university landscape. The case is well suited for such an exploration for several reasons: German universities have a strong publication output and the landscape is diverse (Agasisti & Pohl, 2012). The 121 German universities do not only differ regarding their size but also with respect to the composition of their disciplinary profile. In addition, the advancement of OA has been supported as a priority of funding organisations as well as various institutions. Last but not least, the nationwide consortium DEAL was created, under which two OA transformative agreements were negotiated (Haucap et al., 2021). To sum up, the case of German universities allows us to study the effects of different mechanisms for the advancement of OA in a diverse university landscape.

The article is organised as follows: In a first step, with reference to governance theory, the adoption of OA in universities is understood as a process of organisational governance that involves different dimensions. These dimensions are applied as heuristics to identify three different types of factors that may influence the OA uptake at German universities (Section 2). Hypotheses about possible effects are formulated in Section 3, followed by a description of the methods in Section 4. In Section 5, descriptive statistics are given and the results of the regression models for the determinants of the adoption of OA are presented. In the discussion section (Section 6), the results of the quantitative studies are complemented with insights of an interview study with OA representatives of German universities that give some context and background information about the factors investigated. In addition, the limitations of the study are discussed. The conclusion (Section 7) summarises the most important results.

## 2. Theory

In order to systematise possible factors for the uptake of OA on the level of institutions, we develop a heuristic inspired by a theoretical concept drawn from higher education studies: the governance equaliser. The governance perspective is interested in patterns of social order as a result of intention-based shaping processes, and the particular approach is rooted in the seminal work of Burton Clark (1979). Clark makes clear that governance within a national system of higher education is not a monolithic or concerted action but involves different types – or 'pathways' in his words – of coordination. For that reason, it is not assumed that governance of higher education is a smooth regulation of the system but a complex relation of different types of coordination making it likely that contradictions and tensions may occur, leading to an amplification of effects or conflicting results. The governance equaliser derives from this starting point and understands governance as a multi-dimensional process in which different forces interact. It distinguishes five governance dimensions with corresponding mechanisms, namely state regulation, stakeholder guidance, academic self-governance, managerial self-governance, and competition.

---

[4]    See Martin-Martin et al., 2018 for an overview; Piwowar et al., 2018; Science-Metrix, 2018; Björk & Korkeamaki, 2020.

- *State regulation* can be understood as the "traditional top-down authority, vested in the state" (de Boer et al., 2007). Examples are the prescription of behaviour under specific circumstances, most prominently, but not only, by legislation.
- *Stakeholder guidance* describes the involvement of important, often non-academic stakeholders via means of the participation of goal setting or advice. An example in the German university landscape is the establishment of university councils in which different stakeholder groups are represented.
- *Academic self-governance* refers to the influence of professional communities within the university system. Institutionalised mechanisms in which academic self-governance is exercised are collegial decision-making and peer review.
- *Managerial self-governance* describes a second type of hierarchical relation but this time within university administrations. It influences internal goal setting, regulation and decision-making within universities.
- *Competition* is the fifth governance dimension that describes rivalry for scarce resources like money, staff, prestige and time. In the case of universities, competition usually does not take place in full-fledged markets but in quasi-market settings.

Within this approach it is assumed that at a given point, higher education governance consists of a specific mixture of the different dimensions involved. In the course of time, the composition of the governance regime may change, which implies that the strength and the characteristics of the five dimensions alter. Studies that make use of the governance equaliser can differ regarding the level of analysis that is addressed. On the macro level, the focus is on the relation between higher education institutions and the state. On the meso level, the emphasis is on decision-making within individual institutions and on the micro-level, on decision-making on the level of chairs (Schimank, 2007). The governance equaliser has proved to be fruitful as a heuristics in different thematic fields like the comparisons of the governance of the university systems of different countries (de Boer et al., 2007), professional education (Schemmann, 2014) and – strongly modified – in the introduction of sustainable development goals in universities (Bauer et al., 2018, Niedlich et al., 2020). Up to now, the governance equaliser has not been applied to the adoption of OA on the level of universities. However, with respect to the German university system, the concept seems to be promising, as at least three of the five governance dimensions have played a role in the past.

The probably least controversial dimension is *academic self-governance,* which finds its expression in institutionalised publication cultures. Publication cultures include a set of field-specific publication media. These are subject to generalised quality attributions of the respective scientific community on which a reputation pyramid of the publication media are based, convictions about standards publishable research has to meet together with mechanisms that apply such criteria to select contributions for publication (e.g., peer review). Another component are field-specific routines or practices on how to deal with scientific information that is made available via different channels (Taubert, 2021). With respect to free digital publishing, publication cultures also comprise attitudes towards different OA-types resulting in differences regarding the OA-affinity between different disciplines and fields (Zhu 2017, Dalton et al., 2020). Regarding the history of OA, academic self-governance was the main driver for the adoption for a long time. At the beginning of the 1990s the high energy physics community started to experiment with digital preprint-servers (Ginsparg, 1994; 2011), which marks the starting point for a repository-based OA, often called green OA. Efforts to provide free electronic journals were undertaken in different scientific fields and go back as early as the 1980s (Suber, 2009).

The second dimension that is relevant for the uptake of OA on the level of institutions is *managerial self-governance*. Like in other fields of digital transformation, in higher education in Germany, managerial self-governance has a specific shape also in the case of OA. Instead of a set of rules or prescriptions that is effectively reinforced by sanctions, the predominant mechanisms of managerial self-governance are much softer in character. In the first place, they consist of the build-up and maintenance of digital infrastructures that are accompanied by the provision of specific services as well as positions that are responsible for them. In other words, they are supportive as they aim to reduce the efforts for scientists making their research OA. With inner-organisational regulations, like, for example, OA policies, universities usually express support for OA. On the infrastructural side, managerial self-governance is institutionalised via repositories for depositing (and aggregating) research, publishing platforms like Open Journal Systems (OJS) that are used for the hosting of OA journals, and publication funds for financing article processing charges (APC) or book processing charges (BPC). Institutionalised regulations do not only include OA policies that encourage scientists to publish their research OA but also rules for the usage of the infrastructure and services like funding requirements for publication funds. Finally, the introduction of OA officers as well as OA websites and training activities can be understood as accompanying instruments of managerial self-governance that foster OA. Such infrastructural fields of action can be attributed to managerial self-governance as the libraries that are part of the administration are responsible for it.

Given that freedom of science in Germany is guaranteed by the constitution (Art. 5 III GG), and given that the publication is protected by this right, there are no mechanisms of strong top-down regulations on the level of the state like, for example, mandates that enforce OA.[5] However, if the understanding of state regulation is widened and softer regulations as well as voluntary agreements within the German institutional landscape are considered, some nation-wide coordination efforts can be identified: first, the German Ministry for Education and Research has published a nation-wide OA policy (BMBF 2018) that expresses their support. In addition, the coalition agreement of the federal government mentions OA as a priority in science policy. Second, funding-organisations, like the Deutsche Forschungsgemeinschaft, have implemented programmes that aim to support OA yielding effects on the higher education institutional landscape.[6] Third, transformative agreements have been introduced, and the probably most impactful contracts are those that were negotiated by large publishing houses and project DEAL[7]. To date, they operate on an 'all-in-principle' of nearly all public research institutions, the contracts can be regarded as a central coordination mechanism that affects the entire German research system.

---

[5] The only mandate in Germany that requires the archiving of full texts of publications on a repository was adopted by the University of Konstanz in 2015. The question whether or not the mandate violates the freedom of science guaranteed by the German constitution is subject of a lawsuit (Hartmann, 2017).

# 3. Research question and hypothesis

This study focuses on OA provided by journals ('gold OA', Suber 2012) in the publication output of German universities. It further differentiates between Full OA and Hybrid OA and asks the question to what extent the three governance dimensions (academic self-governance, managerial self-governance and state regulation) affect the OA share of the German university landscape.[8]

- *Full OA (F)*: Full OA means articles published in journals in which all content is immediately freely available online without charging any fees for accessing it.
- *Hybrid OA (H)*: In the case of Hybrid OA, individual articles are made openly available under an open content licence usually by paying a fee, while the journal as a whole remains subscription-access (Jahn et al. 2022).

Full OA and Hybrid OA follow different logics. Therefore, different hypotheses can be formulated:

Full OA

*F-1 Hypothesis on infrastructural requirements (part of managerial self-governance)*

$H_1$: Universities with a publication fund have a larger Full OA share than universities without a publication fund.
$H_0$: Universities with a publication fund have a smaller (or equal) OA share than universities without a publication fund.

*F-2: Hypothesis on the impact of OA policies, OA officers, OA information and OA-events (part of managerial self-governance)*

$H_1$: There is a need for educating researchers regarding the advantages of the APC-liable Full OA model. Universities that are highly engaged in educating and encouraging scientists (provision of OA information and organisation of OA events, supported by OA officers and OA policies) have a higher Full OA share than universities that are less engaged.

---

[6]   The programme "Open-Access-Publizieren" (Ploder et al., 2020) supported the build-up of publication funds on the local level. The programme "Open-Access-Publikationskosten" provides financial support for universities that are engaged in a transformation towards publication-volume based costs (Deutsche Forschungsgemeinschaft, 2022). Both programmes do not aim to directly influence publication behaviour of scientists but fund the build-up of infrastructures that support OA and are therefore linked to managerial self-governance.

[7]   https://www.projekt-deal.de/about-deal/ (accessed August, 29th 2022).

[8]   It is complemented by a second article that focuses on Suber's (2012) other major OA-type, i.e. repository provided OA ('green OA').

$H_0$: Universities that are highly engaged in educating and encouraging scientists (provision of OA information and organisation of OA events, supported by OA officers and OA policies) have a smaller (or equal) Full OA share than universities that are less engaged.

*F-3: Hypothesis on the influence of the disciplinary profile (academic self-governance)*

$H_1$: Universities with a disciplinary profile that shows a strong affinity towards Full OA have a larger Full OA share than universities with a weaker affinity towards Full OA.
$H_0$: Universities with a disciplinary profile that shows a strong affinity towards Full OA have a smaller (or equal) Full OA share than universities with a weaker affinity towards Full OA.

*F-4: Hypothesis on the influence of transformative agreements (DEAL, part of state regulation)*

$H_1$: The larger the share of Full OA articles covered by DEAL, the larger is the overall Full OA share of a university.
$H_0$: The larger the share of Full OA articles covered by DEAL, the smaller is the overall Full OA share of a university.

## Hybrid OA

*H-1 Hypothesis on infrastructural requirements (part of managerial self-governance)*

$H_1$: Universities with a publication fund have a higher Hybrid OA share than universities without a publication fund.
$H_0$: Universities with a publication fund have a smaller (or equal) Hybrid OA share than universities without a publication fund.

*H-2: Hypothesis on the impact of OA policies, OA officers, OA information and OA-events (part of managerial self-governance)*

$H_1$: There is a need for educating scientists regarding the advantages of the APC-liable Hybrid OA model. The stronger a university is engaged in educating and encouraging scientists (provision of OA information and organisation of OA events, supported by OA officers and OA policies) the larger is the Hybrid OA share.
$H_0$: The stronger a university is engaged in educating scientists in OA (provision of OA information and organisation of OA events, supported by OA officers) the smaller is the Hybrid OA share.

*H-3: Hypothesis on the influence of the disciplinary profile (academic self-governance)*

$H_1$: Universities with a disciplinary profile that shows an affinity towards Hybrid OA have a larger Hybrid OA share than universities with a disciplinary profile with less affinity towards Hybrid OA.
$H_0$: Universities with a disciplinary profile that shows an affinity towards Hybrid OA have a smaller (or equal) Hybrid OA share than universities with a disciplinary profile with less affinity towards Hybrid OA.

*H-4: Hypothesis on the influence of transformative agreements (DEAL, part of the state regulation)*

$H_1$: The larger the share of Hybrid OA articles covered by DEAL, the larger is the overall Hybrid OA share of a university.
$H_0$: The larger the share of Hybrid OA articles covered by DEAL, the smaller is the overall Hybrid OA share of a university.

# 4. Data and methods

The study combines three types of data: Bibliometric data of the publication output complemented with OA evidence, structural data and information about OA infrastructures, and interviews with OA officers and OA representatives from German universities.

The *publication output* of German universities was identified through the Web of Science in-house database maintained by the German Competence Center for Bibliometrics (WoS-KB) in its 2021 version. The main advantage of using this data source in the context of our study is that it provides disambiguated address information (Rimmert et al., 2017), which allows obtaining the publication output represented in the Web of Science on the level of institutions with a "near-complete national-scale coverage" of Germany's institutions at a high accuracy (Donner et al., 2020). Publishing articles as Hybrid OA usually involves the obligation to pay APCs, and the same is true for the majority of articles published in Full OA journals as well (Smith et al., 2021). Given this and the fact that the existence of funding mechanisms (like publication funds or transformative agreements) may play a role in the uptake of OA, the analysis was restricted to corresponding author publications. All publications of the period 2010-2020 with a corresponding author from a German university were considered. To identify articles in full OA journals, the ISSN-GOLD-OA 4.0 list (Bruns et al., 2020) and Unpaywall's Full OA journal list were used (Piwowar et al., 2019). The Hybrid OA information was generated as follows: Based on article-level evidence from Unpaywall, articles were assigned to the category Hybrid OA if they were classified as Hybrid OA by Unpaywall and the corresponding journal was not included in the ISSN-GOLD-OA 4.0 list (Bruns et al., 2020). Since preparation of our earlier study investigating OA shares of German research institutions (Hobert et al., 2021), Unpaywall's classification of Hybrid OA has strongly improved (Piwowar et al. 2019). Therefore, we use the distinguished Hybrid OA category instead of our own previous 'other_oa_journal' category, which aside from Hybrid OA included other journal-based OA in not full OA journals (like Moving Wall OA or openly available articles on the publisher's webpage without any open licence).

*Academic self-governance* was conceptualised by one factor on a high level of aggregation. For each of the 255 WoS subject categories, a subject and OA category-specific share was calculated based on all publications with a corresponding author from a German institution. Based on the subject category OA shares and the number of publications in each subject category, a disciplinary influence factor was calculated for all universities and for both Full OA ($X_i^F$) and Hybrid OA ($X_i^H$), namely

$$X_1^F(i) \;=\; \frac{1}{T_i}\sum_{s\,\in\,S}\;\left(N_{i,s} * P_s^F\right)\text{, and } X_1^H(i) \;=\; \frac{1}{T_i}\sum_{s\,\in\,S}\;\left(N_{i,s} * P_s^H\right)$$

where

$X_1^F(i)$    Full OA disciplinary influence factor for university $i \in I$ the set of all included universities,

$X_1^H(i)$    Hybrid OA disciplinary influence factor for university $i \in I$,

$N_{i,s}$    Number of publications of university $i \in I$ in WoS subject category $s \in S$ the set of all WoS subject categories,

$T_i$    Total number of publications of university $i \in I$,

$P_s^F$         Full OA share of WoS subject category $s \in S$ (publications with German corresponding authors), and

$P_s^H$         Hybrid OA share of WoS subject category $s \in S$ (publications with German corresponding authors).

For *managerial self-governance,* a structured data collection was created[9] by harvesting different sources of information and by manual online search. The data set includes information about the size of universities (in terms of students, staff, professors, budget, and third-party funds[10]) as well as OA infrastructures and services that are provided on the local level. The last mentioned data include information about the existence of publication funds[11], OA policies[12], OA officers[13], OA websites and OA activities like information events or workshops announced on the universities' websites.[14] Data collection took place between August and October 2021. The data are modelled as response variables $X_2$ to $X_8$.

A number of mechanisms of *state regulation* set the same framework conditions for the entire German university landscape. Given that these aspects do not vary between institutions, they cannot be used to explain differences in the adoption of OA. However, data is available for one of the arguably most important mechanisms of state regulation today, namely the transformative agreements negotiated between project DEAL and the large publishing houses Wiley and Springer. Although these contracts operate on an all-in principle and include all German universities, the number of publications covered by the two DEAL contracts vary from university to university as their publication output in journals covered by the contracts differ. For our analysis of a possible influence of the DEAL contracts, the publication year 2020 is considered as this is the only year for which the transformative agreements with Springer and Wiley both have been effective for the whole year and for which data are available.[15] For each university and for each response variable (Full and Hybrid OA shares) we calculated the share of the publication output covered by DEAL contracts as

$$X_9^F(i) = \frac{DF_{(i)}}{TP_{(i)}}, \text{ and } X_9^H(i) = \frac{DO_{(i)}}{TP_{(i)}}$$

where

$X_9^F(i)$    Share of Full OA publications covered by DEAL contracts for university $i \in I$, the set of all included universities,

$X_9^H(i)$    Share of Hybrid OA publications covered by DEAL contracts for university $i \in I$,

$D_i^F$       Number of Full OA publications covered by DEAL contracts for university $i \in I$,

$D_i^H$       Number of Hybrid OA publications covered by DEAL contracts for university $i \in I$, and

$T_i$        Total number of publications of university $i \in I$.

---

Table 1 gives an overview of the explanatory and response variables that are considered in the regression models together with their labels.

**Table 1: Explanatory and response variables**

| Governance dimension | Variable description | Label |
|---|---|---|
| Academic self-governance | Estimated Full OA share / Hybrid OA share based on the composition of subjects | $X_1^F / X_1^H$ |
| Managerial self-governance | Existence of an OA publication fund | $X_2$ |
| | Existence of an OA officer | $X_3$ |
| | Existence of a webpage with OA information | $X_4$ |
| | Existence of a webpage with information about OA activities | $X_5$ |
| | Existence of a webpage with OA rights information | $X_6$ |
| | Existence of an OA policy | $X_7$ |
| | Month of OA policy adoption | $X_8$ |
| State regulation | Share of Full OA / Hybrid OA journal articles covered by DEAL contracts | $X_9^F / X_9^H$ |
| Response variable | Full OA share / Hybrid OA share in the publication output of German universities | $Y^F / Y^H$ |

In order to put our statistical model into a broader context and to gain more detailed insights into how the different aspects of the three governance dimensions influence the OA shares of German universities, expert interviews with OA officers and representatives from 20 different universities were conducted between February and June 2021. The selection of interviewees aims to represent a large diversity of perspectives and follows the selection scheme of maximum variation (Collins et al., 2006, 84). It includes interviewees from large and small universities, universities with strong or weaker OA adaption as well as universities with different disciplinary profiles (with and without a medical faculty, technical universities and universities with a broad disciplinary mixture). The interview guideline covers all governance dimensions and factors that are included in the regression models and the duration of the interviews varied between 47 to 119 minutes. All interviews were transcribed, coded and analysed with content analysis (Mayring, 2015) using MAXQDA 2018 as data analysis software. Evidence from the interviews will be used to contextualise the results of the regression models in the discussion section.

# 5. Results

## 5.1 Descriptive statistics

In a first step, descriptive statistics are reported for categorical and metrical explanatory and response variables. However, the availability of data differs. Structural information about the German university landscape and about OA infrastructures were collected at a specific point in time when the manual research took place. In contrast, publication-based information like publication output, OA shares, and disciplinary influence scores can be calculated from data spanning different periods. Finally, information about publications covered by DEAL contracts is available for the publication year 2020. With the exception of DEAL-shares, publication-based indicators are given for three periods (2010-2020, 2017-2018, and 2020) for which regression models are calculated. The rationale for the selection of the three periods is to analyse and compare the influence of the three governance dimensions for the whole 11-year period, with the most recent period before the introduction of the DEAL contracts (2017-2018) and the period for which information about the DEAL contracts are available (2020).

Table 2 gives an overview of the descriptive statistics for categorical independent variables and illustrates that German universities differ regarding the mechanisms and activities they have implemented to support OA. While more than 80 % of the universities have a website with OA information, nearly three quarter have an OA officer, 70 % a publication fund and nearly two thirds an OA policy. Only half of the universities provide OA rights information and a bit more than a third of them information about OA courses and training on their websites.

**Table 2: Descriptive statistics for categorical independent variables**

| Variable | True | True (%) | False | False (%) |
|---|---|---|---|---|
| $X_2$ (publication fund) | 73 | 70.2 | 31 | 29.8 |
| $X_3$ (OA officer) | 77 | 74.0 | 27 | 26.0 |
| $X_4$ (webpage with OA information) | 87 | 83.7 | 17 | 16.4 |
| $X_5$ (OA activities) | 37 | 35.6 | 67 | 64.4 |
| $X_6$ (OA rights information) | 51 | 49.0 | 53 | 51.0 |
| $X_7$ (OA policy) | 67 | 64.4 | 37 | 35.6 |

Descriptive statistics for the duration of the adoption of OA policies at German universities are given in table 3. The first line includes both universities with as well as without OA policies. For universities without OA policy, the duration of policy adoption was defined as 0 months. The statistics in the second line are limited to universities with OA policies.

**Table 3: Descriptive statistics for metrical independent variables**

| Variable | Observations | Mean | Std. Dev. | Min | Max |
|---|---|---|---|---|---|
| $X_8$ (months of policy adoption, all universities) | 104 | 48.77 | 53.02 | 0 | 179 |
| $X_8$ (months of policy adoption, universities with OA policy) | 67 | 75.70 | 48.14 | 0 | 179 |

The publication-based descriptive statistics are presented in table 4. The table includes descriptive statistics for the total number of publications, Full and Hybrid OA share as well as the Full and Hybrid OA disciplinary influence factors for all periods. For the publication year 2020, DEAL influence factors were calculated both for Full and Hybrid OA. In a first step, all indicators were calculated for each university that overrun the threshold value of a publication output of 50 corresponding author publications for the particular period. The threshold value was introduced to exclude distortions of the OA shares and disciplinary influence factors due to small publication output. In a second step, mean value and standard deviation were calculated and minimum and maximum values were given for the German university landscape. The results in the table show that all OA shares have increased for more recent years. Particularly noteworthy is the rise of the Hybrid OA share between the period 2017-2018 and 2020 from 4.6 % to 21.8 % with a maximum of 30.7 % for one university.

**Table 4: Publication-based indicators (independent and dependent variables)**

| Variable | Observations | Mean | Std. Dev. | Min | Max |
|---|---|---|---|---|---|
| **2010-2020** | | | | | |
| Total Publications* | 98 | 6,203.48 | 6,530.22 | 54 | 26,912 |
| $Y^F$, Full OA (%)* | 98 | 14.48 | 5.48 | 0.63 | 27.38 |
| $Y^H$, Hybrid OA (%)* | 98 | 5.71 | 1.80 | 1.32 | 11.46 |
| $X_1^F$, Full OA disciplinary influence factor* | 98 | 12.46 | 3.69 | 1.76 | 20.27 |
| $X_1^H$, Hybrid OA disciplinary influence factor* | 98 | 6.90 | 0.79 | 4.64 | 8.24 |
| **2017-2018** | | | | | |
| Total Publications | 82 | 1,481.74 | 1,310.11 | 59 | 5,366 |
| $Y^F$, Full OA (%)** | 82 | 18.39 | 6.00 | 0 | 34.79 |
| $Y^H$, Hybrid OA (%)** | 82 | 4.58 | 1.84 | 0 | 8.64 |
| $X_1^F$, Full OA disciplinary influence factor** | 82 | 15.47 | 4.37 | 2.08 | 25.69 |
| $X_1^H$, Hybrid OA disciplinary influence factor** | 82 | 6.34 | 0.79 | 4.19 | 7.44 |
| **2020** | | | | | |
| Total Publications | 79 | 876.95 | 745.69 | 55 | 3,126 |
| $Y^F$, Full OA (%)*** | 79 | 26.20 | 8.19 | 7.94 | 58.15 |
| $Y^H$, Hybrid OA (%)*** | 79 | 21.80 | 3.79 | 8.82 | 30.74 |
| $X_1^F$, Full OA disciplinary influence factor*** | 79 | 22.57 | 5.45 | 8.02 | 39.55 |
| $X_1^H$, Hybrid OA disciplinary influence factor*** | 79 | 24.55 | 1.51 | 18.99 | 27.60 |
| $X_9^F$, share of Full OA DEAL publications*** | 79 | 1.93 | 1.11 | 0 | 5.32 |
| $X_9^F$, share of Hybrid OA DEAL publications*** | 79 | 9.83 | 3.25 | 0 | 15.30 |

*Universities with a publication output > 50 in 2010-2020, restricted to publications with corresponding authors of that university
** Universities with a publication output > 50 in 2017-2018, restricted to publications with corresponding authors of that university
*** Universities with a publication output > 50 in 2020, restricted to publications with corresponding authors of that university

## 5.2 Regression models

Multiple linear regression analysis is an important statistical tool to test assumptions about structures and relations in data (Freedman, 2009). In regression analysis, the output variable is named dependent variable, and the variables that are assumed to have effects on the dependent variable are called independent variables. In our analysis, separate regression models were calculated for three time periods and two dependent variables each (Full OA share and Hybrid OA share). Given that collinearity of explanatory variables can be a problem for regression analysis, variance inflation factors (VIFs) were calculated for all regression models using the STATA 11 *VIF* function.

**Table 5: Variance inflation factors**

| | $X_1^F$ | $X_1^H$ | $X_2$ | $X_3$ | $X_4$ | $X_5$ | $X_6$ | $X_7$ | $X_8$ | $X_9^F$ | $X_9^H$ |
|---|---|---|---|---|---|---|---|---|---|---|---|
| **Full OA** | | | | | | | | | | | |
| 2010-2020 | 1.23 | – | 1.71 | 1.70 | 2.19 | 1.25 | 1.44 | 2.77 | 2.10 | – | – |
| 2017-2018 | 1.18 | -- | 1.50 | 1.55 | 1.99 | 1.21 | 1.34 | 2.41 | 1.93 | | |
| 2020 | 3.50 | – | 1.40 | 1.43 | 1.58 | 1.21 | 1.22 | 1.99 | 1.80 | 3.40 | – |
| **Hybrid OA** | | | | | | | | | | | |
| 2010-2020 | | 1.60 | 1.70 | 1.66 | 2.27 | 1.26 | 1.42 | 2.71 | 2.27 | -- | -- |
| 2017-2018 | – | 1.41 | 1.48 | 1.61 | 1.73 | 1.20 | 1.21 | 2.05 | 1.96 | – | – |
| 2020 | -- | 1.29 | 1.39 | 1.40 | 1.63 | 1.18 | 1.26 | 1.96 | 1.91 | – | 1.27 |

The values in table 5 show that there is some explanatory power between the independent variables but they all are well below the critical value of 5, which is considered as a threshold value above which the model should be adjusted, e.g. by excluding certain independent variables. As a consequence, all considered variables are included in the regression analysis.

**Table 6: Full OA, regression models**

| Reg. | $X_1^f$ | $X_2$ | $X_3$ | $X_4$ | $X_5$ | $X_6$ | $X_7$ | $X_8$ | $X_9^f$ | F | $R^2$ | Adj. $R^2$ | RMSE |
|---|---|---|---|---|---|---|---|---|---|---|---|---|---|
| **2010-2020** | | | | | | | | | | | | | |
| 1 | 1.349** | 0.020** | -0.004 | -0.036** | 0.006 | -0.001 | 0.007 | 0.000 | -- | 68.07 | 0.860 | 0.847 | 0.021 |
| 2 | 1.306** | 0.014* | -- | -- | -- | -- | -- | -- | -- | 216.60 | 0.820 | 0.816 | 0.023 |
| 3 | 1.338** | -- | -- | -- | -- | -- | -- | -- | -- | 405.04 | 0.808 | 0.806 | 0.024 |
| **2017-2018** | | | | | | | | | | | | | |
| 4 | 1.297** | 0.018* | -0.002* | -0.026* | 0.004 | -0.006 | 0.004 | 0.000 | -- | 78.44 | 0.888 | 0.877 | 0.023 |
| 5 | 1.288** | 0.016* | -- | -- | -- | -- | -- | -- | -- | 279.81 | 0.868 | 0.865 | 0.024 |
| 6 | 1.317** | -- | -- | -- | -- | -- | -- | -- | -- | 523.03 | 0.859 | 0.857 | 0.024 |
| **2020** | | | | | | | | | | | | | |
| 7 | 1.055** | 0.039** | 0.010 | -0.072* | -0.001 | -0.020 | 0.021 | -0.000 | 1.597* | 32.39 | 0.809 | 0.784 | 0.038 |
| 8 | 1.297** | 0.037** | -- | -- | -- | -- | -- | -- | -- | 125.97 | 0.768 | 0.762 | 0.040 |
| 9 | 1.016** | -- | -- | -- | -- | -- | -- | -- | 1.663* | 118.92 | 0.758 | 0.752 | 0.041 |
| 10 | 1.295** | -- | -- | -- | -- | -- | -- | -- | -- | 220.57 | 0.741 | 0.738 | 0.420 |

* Significant at the 0.05 level; ** significant at the 0.01 level

**Table 7: Hybrid OA, regression models**

| Reg. | $X_1^H$ | $X_2$ | $X_3$ | $X_4$ | $X_5$ | $X_6$ | $X_7$ | $X_8$ | $X_9^H$ | F | $R^2$ | Adj. $R^2$ | RMSE |
|---|---|---|---|---|---|---|---|---|---|---|---|---|---|
| **2010-2020** | | | | | | | | | | | | | |
| 11 | 1.162** | 0.004 | 0.007 | -0.003 | 0.001 | 0.001 | -0.003 | 0.000 | -- | 8.61 | 0.436 | 0.386 | 0.014 |
| 12 | 1.434** | -- | -- | -- | -- | -- | -- | -- | -- | 62.49 | 0.394 | 0.388 | 0.014 |
| **2017-2018** | | | | | | | | | | | | | |
| 13 | 1.480** | 0.002 | 0.000 | 0.007 | 0.003 | -0.000 | 0.002 | -0.000 | -- | 7.84 | 0.462 | 0.403 | 0.014 |
| 14 | 1.550** | -- | -- | -- | -- | -- | -- | -- | -- | 63.81 | 0.444 | 0.437 | 0.014 |
| **2020** | | | | | | | | | | | | | |
| 15 | 0.977** | -0.002 | -0.013 | 0.040 | -0.001 | -0.000 | -0.003 | 0.000 | 0.418** | 6.07 | 0.442 | 0.369 | 0.030 |
| 16 | 1.043** | -- | -- | -- | -- | -- | -- | -- | 0.439** | 27.07 | 0.416 | 0.401 | 0.029 |
| 17 | 1.356** | -- | -- | -- | -- | -- | -- | -- | -- | 31.50 | 0.290 | 0.281 | 0.032 |

* Significant at the 0.05 level; ** significant at the 0.01 level

*Full OA*

To begin with the most definite result for Full OA, namely hypothesis F-3 on the influence of the disciplinary profile, $H_O$ has to be rejected as all simple linear regression models (nos. 3, 6 and 10) show strong effects of the disciplinary influence factor $X_1^F$ on the Full OA share of German universities, and the regression coefficient for the factor can be interpreted as follows: Depending on the period that is considered, the Full OA share of a university raises between 1.295 and 1.338 percent points if the disciplinary influence factor increases by one percent point. The composition of the disciplinary profile is by far the most important variable that alone explains from 73.8 % (2020) to 85.7 % (2017-2018) of the variance of the dependent variable as the coefficient of determination adj. $R^2$ of the univariate regression models shows. The strong effect of the disciplinary factor remains even when we control for other possible influences in multiple regression analyses (nos. 1, 4 and 7) including all of the independent variables described before in Section 4 (with the exception of the DEAL-related factor, which is discussed below). Therefore, academic self-governance is the most determining factor for the uptake of Full OA at German universities.

Regarding Hypothesis F-1, $H_O$ ('Universities with a publication fund have a smaller (or equal) OA share than universities without a publication fund') is rejected for all periods on the 0.05 level of significance and for 2010-2020 and 2020 also on the 0.01 level of significance in the full models. For all years, the existence of a publication fund turns out to have a small but significant positive effect on the Full OA share. A comparison of the models including the disciplinary influence factor $X_1^F$ and the existence of a publication fund $X_2$ with the models where the disciplinary influence score $X_1^F$ is the only independent variable, the inclusion of the variable $X_2$ adds only little explained variance with the strongest effect in the period 2020 (77.1 % of the variance explained compared to 73.8 % using only the disciplinary factor).

With respect to F-2, $H_O$ ('Universities that are highly engaged in educating scientists (provision of OA information, organisation of OA events, support of OA officers and OA policies) have a smaller or equal Full OA share than universities with less engagement) cannot be rejected. In all three periods, $X_4$ (webpage with OA information) turned out to be significant at least at 0.05 level, but the effect points in the direction of $H_O$. The same holds for $X_3$ in the period 2017-2018. In other words, no significant positive effect of the existence of an OA officer, OA webpage, OA rights information and OA training activities on the Full OA share could be established by the regression analysis. When the results of the regression analyses nos. 1, 4 and 7 are compared with the univariate regression analyses nos. 3, 6 and 10, it turns out that the inclusion of the variables of managerial self-governance shows only small improvements of the explained variance, represented (adj. $R^2$ value of 0.847 vs. 0.806, 0.877 vs. 0.857, and 0.784 vs. 0.738).

Finally, hypothesis F-4 formulates a conjecture about the influence of transformative agreements, namely the large contracts with SpringerNature and Wiley that were negotiated by project DEAL. These contracts became effective in 2019 and 2020, respectively. The regression analysis provides evidence that $H_O$ (Universities with a large share of Full OA-articles covered by DEAL transformative agreements, have an equal or a smaller Full OA share than universities with a smaller Full OA share in transformative agreements) has to be rejected and that DEAL has a positive effect on the Full OA share on a 0.05 level of significance when controlling for other factors. However, the added explained variance of the Full OA share of universities by the share of publications covered by DEAL is relatively small (additional 1.4 % explained variance).

*Hybrid OA*

Turning to Hybrid OA the results of the regression models are in accordance with those for Full OA but also differ in part. To begin again with hypothesis H-3, $H_O$ ('Universities with a disciplinary profile that shows affinity towards Hybrid OA have a smaller or equal Hybrid OA share than universities with a disciplinary profile with less affinity towards Hybrid OA') can be rejected. According to adj. $R^2$, the proportion of the variance explained by the disciplinary influence factor is reasonably large and varies between 28,1% and 43,7%. However, when compared with the Full OA models, it ranges on a much lower level. The regression coefficient shows that the Hybrid OA share increases between 1.356 (period 2010-2020) and 1.550 (2017-2018) percent points if the disciplinary influence score increases by one point.

Regarding the effect of managerial self-governance, all variables ($X_2$ to $X_8$) turned out not to be significant in any of the periods analysed. Therefore, our data support neither H-1 ('Universities with a publication fund have a higher Hybrid OA share than universities without a publication fund') nor H-2 ('There is a need for educating scientists regarding the advantages of the APC-liable Hybrid OA model. The stronger a university is engaged in educating scientists (provision of OA information and organisation of OA events, supported by OA officers and OA policies) the larger is the Hybrid OA share').

Finally, the effect of the DEAL transformative agreements was considered for the year 2020. When combined with the disciplinary influence factor $X_1^H$, the (at 0.01 level) significant DEAL influence factor $X_9^H$ adds 12,0% of explained variance to the model. Hypothesis H-4 ('The larger the share of Hybrid OA articles covered by DEAL, the larger is the overall Hybrid OA share of a university') is therefore supported. In addition, it is interesting to note that the additional explained variance of the DEAL influence factor is much higher in the case of Hybrid OA than in the case of Full OA.

# 6. Discussion

In this section, we will summarise the results of the regression analysis and discuss them in the context of interviews that were conducted with OA officers at German universities to deepen the understanding of the uptake of OA.

The most important results can be summarised as follows: Regarding the adoption of both Hybrid and Full OA, the most determining factor is the disciplinary profile of German universities. The more a university's publication output comprises publications from subject fields with a high degree of (Hybrid or Full) OA adaption, the larger is the (Hybrid or Full) OA share of the university. With reference to the different governance dimensions introduced in the theoretical section, academic self-governance is the driver of the adoption of journal-based OA. In the interpretation of the results it should be kept in mind that the analysis happens on a high level of aggregation, resulting in at least two limitations: The first one is a consequence of the way in which the disciplinary influence factor is conceptualised. As a highly aggregated factor, it reflects the adoption in all fields of science by a single number. Therefore, it is not possible to attribute differences in the adoption of OA to individual scientific areas, disciplines, specialties or fields. Second, the regressions show that the disciplinary profile is by far the strongest determinant but the actual mechanisms of how the disciplinary publication culture affects the OA share remains unclear. The analysis cannot answer the question whether it is the attitude of scientists in fields with an affinity towards OA, the existence of practices and routines in the context with OA publication media (Taubert, 2021), the availability of Hybrid or Full OA journals (Severin et al., 2020), or a combination of two or more factors that is decisive here.

Regarding state regulation, the influence of transformative agreements that were negotiated in the context of project DEAL could be tested for the year 2020 with mixed results. For both OA types, transformative agreements turn out to be a significant factor but the explanatory power differs. In the case of Full OA the factor adds only a small fraction of explained variance compared to a model using just the disciplinary influence factor, while in the case of Hybrid OA the explanatory power of the model is substantially improved. For Hybrid OA, such agreements can be considered as an effective instrument, yielding remarkable results on the level of the university landscape.

Besides the strong effects of academic self-governance and state regulation (the latter mainly in the case of Hybrid OA), the weak explanatory power of the variables of managerial self-governance is another noticeable result. Most of the variables of managerial self-governance are not significant in any regression model and those that are significant point in the direction of $H_0$ or add only tiny fractions to the explained variance. This result should not be interpreted in the way that the build-up of OA infrastructures, staff and services is no effective means to support OA at individual universities. A number of universities could achieve large journal-based OA shares by the provision of good infrastructures and services. However, when analysed on the level of the whole German university landscape, the factors do not explain much. For a more detailed understanding, the different variables of managerial self-governance are discussed and contextualized using the conducted interviews with OA officers in the following.

*Publication fund ($X_2$)*

To begin with the existence of a publication fund ($X_2$), the variable does not have a significant influence on the Hybrid OA share in any of the periods, but it has small explanatory power for the Full OA share in 2020. To better understand this result, it is worth noticing that the build-up of publication funds in Germany was strongly influenced by the programme 'Open Access Publizieren' of the Deutsche Forschungsgemeinschaft. It supported the build-up of 57 publication funds at German universities (Ploder et al., 2020, p.14) and aimed to establish structures at universities that organise the financial flows and monitor the costs for OA. The DFG programme defined criteria for the financial support including a price-cap for APCs charged by Full OA journals and excluding articles in hybrid journals from funding (Ploder et al., 2020, p. 17). The latter criterion explains why the variable has no explanatory power for Hybrid OA shares of German universities. However, for an understanding of a weak or missing explanatory power in the case of Full OA, different aspects have to be considered. First, a number of interviewees confirm that the DFG criteria that were implemented during the build-up of a publication fund are often still applied after DFG funding has expired.[16] One interviewee describes the influence of the DFG programme beyond the funding period as follows:

> "This year, we are now in the situation to finance our publication fund independently and could define the criteria for funding ourselves. But we did not. We still have the same funding criteria like in 2020 the last year of the DFG-funding of our publication fund. Therefore, the DFG still has a big influence that I find positive. First, a clear commitment to support Gold OA but not Hybrid OA and a price cap of 2,000€ as a maximum financial support" (I-12, pos. 47).

The price-cap of 2,000€ may explain that the effect of publication funds is limited as publications in Full OA journals with higher APCs are excluded from support by the publication fund. The restriction may also continue to be effective after DFG funding has expired. However, the sheer number of 12,000 articles that received funding from DFG sources between 2011-2017 (Ploder et al., 2020, p.33) would suggest that the effect on the Full OA share of Germans university landscape would be larger. Hints for an explanation of the limited effects can be derived from the literature. A retrospective analysis of the outcomes of the already mentioned DFG-programme reports similar Full OA shares for the two groups of universities with and without a DFG publication fund (Ploder et al., 2020, p. 42). In addition, in an analysis of the coverage of APC-liable publications of a university in their publication funds, Bruns & Taubert (2021) found out that a considerable part of such payments - varying between 10.4 % and 89.0 % - were not processed by publication funds but via other channels. Both findings suggest that publication funds are not primarily a financial source that allows scientists to turn additional articles in APC-liable journals OA but, when introduced, are used by scientists to substitute other sources for payments like third party funds or budgets of faculties instead of turning additional articles into OA. This interpretation is also supported by an interviewee who was complemented for the build-up of the publication fund at his university and doubts that the raise of the OA share is an effect of his efforts:

> "Sometimes I ask myself, if this [the growth of the OA share] is my merit. At the time when our former director retired, he said that I had built up the fund and it is incredible how it is being used by now. I ask myself, if there is actually growth within certain disciplines because of the fund or because of their publication culture. Well, from the beginning there was no need to convince them of Open Access. When we built-up the publication fund, they were the first that used it, and this continued" (I-16, pos. 25).

*Open access policy ($X_7$) and month of policy adoption ($X_8$)*

---

[16]  I-01, pos. 19; I-05, pos. 81; I-06, pos. 111, 121; I-07, pos. 60-61; I-11- pos. 45; I-12, pos. 21, 47; I-16, pos. 19; I-17, pos. 77, I-18, pos. 67.

Another instrument for the advancement of OA are institutional OA policies modelled as two variables ('existence of an OA policy' ($X_7$), 'duration of policy adoption' ($X_8$)) in the regression analyses. In all regression models for Full and Hybrid OA and for all periods both variables turned out to be non-significant on a 0.05-level. At first glance, these findings contradict studies on a global scale that report high compliance rates of OA policies and mandates also for journal-based OA (Gargouri et al., 2012, Larivière & Sugimoto, 2018, Kirkmann & Haddow, 2020). However, OA policies and mandates vary in strength (Vincent-Lamarre et al., 2016) as OA can either be 'requested' or 'required'. In addition, non-compliance can be but not always is linked to sanctions like the suspending of payments in the case of funder mandates (Larivière & Sugimoto, 2018) or the non-consideration of non-OA publications in research evaluations like, for example, in the case of the institutional OA policy of the University of Liège (Rentier & Thirion, 2011). Both mandate strength and sanctions would support high compliance rates.

Against this background it is less surprising that the OA policies at German universities do not yield strong effects since institutional OA policies in Germany so far do not formulate mandatory requirements but 'recommend' and 'encourage' to publish OA with only one known exception. The reason for such soft-style policies is that the German constitution guarantees freedom of research, including the freedom of publication. The results of the bibliometric analysis find their reflection in the statements of OA representatives from the interviews. Nearly all of the interviewees do not see any direct effect of OA policies on the journal-based OA share of their universities, and a number of them explicitly reject such a relation.[17] One example of such a perspective can be found in the interview with I-10 who describes the effects of the introduction of the OA policies at her university as follows:

> "I would say, it [the OA policy] has minor or no effects. At the time when the policy was new – and the same also holds for the research data policy that we have – I received a number of nervous telephone calls [from scientists]. "We now have to publish OA, how can we do that?" When I explained during the conversation that the character is more a recommendation than a requirement, the callers quickly took leave" (I-10, pos. 41).

Given that OA policies do not improve the journal-based OA shares and given that OA officers do not expect such effects, the question arises for what purpose German universities have established those documents. Again, insights from the interviews are helpful to understand the underlying factors and mechanisms: No less than 12 interviewees[18] of the 16 universities with an OA policy from the interview sample reported that the trigger for establishing a policy was the DFG project 'Open Access Publizieren'. Within the programme, the existence of an OA policy was not a formal requirement but it was regarded as being beneficial for the proposals by the applicants. One OA officers portrays her argument that was convincing for the implementation of an OA policy at her university as follows:

> "Well, I have said, the DFG wants that [an OA policy] and if the DFG wants that, the scientists want that as well. One wants to have a good standing at the DFG and therefore, I

---

said that it's no drawback if we have such a thing. Because of this reason, it passed [the committees of the university] without resistance" (I-11, pos. 37).

The emergence of many OA policies shows that it responded to external expectations and a demand for legitimation that should increase the chances for the acquisition of resources for the university. However, external legitimation is not the only function OA policies have. It is also a means for legitimising the goal of OA to research internally, as a number of quotations show.[19]

> "I think it is something that the university takes a stand for. This is important for our argument, if we try to promote or support open access and the different publication models of open access […] But it helps to have such a policy on the level of the university, that was adapted with consensus, to rely on that and to create momentum" (I-08, pos. 54).

To summarise, evidence from the regression analysis and the interviews underlines that OA policies of German universities do not have a strong direct effect on the journal-based OA publication output. However, their function is more subtle as they help to legitimise the goal of OA internally and position the libraries as being responsible for the provision of services and for the advancement of OA. In addition, OA policies are important means for an external legitimation that respond to expectations of funders and help to support the flow of resources.

*OA officer ($X_3$)*

In management literature, it is stated that an "important element" for organisational change is "represented by a member of staff or delegate [...] serving the need of a clear structure and continuity. This particular stakeholder should be positioned closely to the senior leadership" (Bauer et al. 2018). In the case of the transformation towards OA, an OA officer could play such a role and may act as a stakeholder in favour of OA. Albeit, the regression analyses do not show any significant effects of the existence of an OA officer in the direction of $H_1$ for Hybrid as well as for Full OA in any of the periods. This result is open to two interpretations. First, individual OA officers may provide information, support and resources to scientists of their institution, but the effect of their efforts of making publications OA is too small to be significant on the level of the German university system. Before accepting this interpretation, one should also consider the possibility of an oversimplified operationalisation of the variable in the regression analysis. Second, the interviews provide evidence that there is a large variance in the way in which the responsibility for OA is incorporated into the role structure of universities. On the one side of the spectrum where an OA officer exists, he or she is the only person who is responsible for OA and related services and tasks. An example is I-19, an OA officer at a large university who describes his position as follows:

> "Actually, I am alone. Well we only have me as an OA representative but not an OA office or something and we do not have any staff in that matter" (I-19, pos. 8).

---

On the other side, there are universities with a highly differentiated role structure and a considerable number of staff, each of them being responsible for different OA services:

> "Virtually, there are three OA-centres at our university. Gold OA in the acquisition section including the transformation budget. The section 'publication and e-learning services' runs not only our repositories, which also includes an image database that is partially OA, but also an OJS-system that is much recommended to the scientists especially in the context of specialised information services (Fachinformationsdienste). And in the field of e-learning we also started a project for the hosting of open educational resources" (I-08, pos. 12).

For the regression analysis, both universities show the same value of the binary variable. In addition, the standing of the OA officer regarding the university leaders may also vary and such differences are also not reflected by the variable. Hence, based on the evidence presented here it cannot be decided if a more differentiated operationalisation of OA professionals at German universities would have resulted in a larger share of explained variance.

*Webpage with OA information ($X_4$), information about OA activities ($X_5$) OA rights information ($X_6$)*

As already shown in the section on descriptive statistics (Section 5.1), German university libraries use a number of channels to inform and train their scientists about OA. These include web pages with OA information ($X_4$), web pages with OA rights information ($X_6$) and other OA activities ($X_5$) like courses, talks and events. Again, none of the variables are significant for either of the OA types and periods with the exception of the existence of webpages with OA information, which is significant but points in the direction of $H_0$.[20] In the case of information provided on websites, the interviews do not help to understand these results: Although addressed by questions in the interview guideline, neither information about OA nor OA rights information on websites were met with much interest of the interviewees. Most of them pointed to the existence of such pages and explained the content but had difficulties in answering the question as to what extent the provided information is used and whether it has effects on OA publishing. One example is interviewee I-21 who comments on the website of her library as follows:

> "For all of our disciplines we have a webpage with discipline-specific information and hints to important databases and so on. And there, we also have the bullet point "Open Access publication in your discipline" so that people who visit the page look into such things. These are probably not so many, but anyway they received a hint on BASE[21] on repositories and so on via that way" (I-21, pos. 87).

This is somewhat in contrast to the passages in which the interviewees are being asked about OA activities like training courses and events, a topic that is discussed at length with diverse perspectives. On the one hand, a number of interviews report (and also complain about) a missing interest of scientists in OA courses and events organised by the library. This is evidenced by the small number of participants that attend such events.[22]

---

[20]  The coefficient of the variable is negative, suggesting that the provision of OA information on a website leads to a smaller Full OA share.

[21]  I-21 refers to the Bielefeld Academic Search Engine (https://www.base-search.net/, accessed August 29th, 2022) here that allows to search for OA publications.

> "That is always the question. How are the talks attended? Usually by a single-digit number. Somehow five to seven participants. We are a small university, though" (I-02, pos. 124).

> "Well it could be more. I would say it is constant, that I always have two or three participants, with the exception of requested seminars, where there are more" (I-04, pos. 75).

Both quotations suggest that OA activities like training and events do not attract many participants and may therefore not have strong effects on the shares of journal-based OA. On the other hand, a contrastive perspective can be found in other interviews that report much interest from scientists and draw a more impactful picture about OA training courses.

> "The [OA] workshops that I teach cannot be provided each month. They are always fully booked and we have to develop the concept a bit because of the limited number of participants that are allowed. [...] There are 35 participants allowed in the workshops because the participants have to show active participation as they receive ECTS credit points for it " (I-06, pos. 7).

> "And in the case of courses for doctoral students there can be 40 plus participants. Courses that are addressed to scientists, I always say that if the number of participants is two-digit, I am fine" (I-08. pos. 92).

A systematic comparison of well and less attended OA courses point to conceptual differences between the two. In the interviews it is reported that courses that are provided proactively by libraries usually yield less attention and smaller audiences than courses, talks and workshops that are delivered upon request[23]. Such requests typically originate from organisational entities within the universities like institutes, faculties, graduate centres and programmes or from academic bodies[24].

The interviews illustrate a large diversity both in the frequency and in the way the courses are conceptualised by university libraries. For the impact of such activities, the interviews suggest that it is decisive if they are part of a proactive teaching programme of libraries or provided on request and if general or subject specific information are provided by them. Therefore, it is possible that the regression analysis might have yielded more meaningful results if the factor 'OA activity' would have been operationalised with a more differentiated and complex set of variables.

---

[22]  I-01 pos. 67, I-04, pos. 75; I-07, pos. 67; I-11, pos. 60; I-16, pos. 81; I-09, pos. 115; I-19, pos. 143; I-21, pos. 81.

[23]  I-02, pos. 122, pos. 124; I-06, pos. 7, pos. 16-19; I-08, pos. pos. 92I-12, pos. 73; I-17, pos. 87; I-20, pos. 11, pos. 99, I-16. pos. 75, 81.

[24]  I-16, pos. 75; I-02, pos. 122.

# 7. Conclusion

This article asked the question as to what factors explain the differences in the uptake of journal-based OA in the German university landscape and distinguished in the analysis between Full OA and Hybrid OA. With respect to theory, the governance equaliser was used as heuristics and three possible explanatory factors were differentiated, namely academic self-governance, managerial self-governance and state regulation. For both OA types the most determining factor for the differences in the OA shares is academic self-governance: The more a university's publication output comprises publications from subject fields with a high degree of (Hybrid or Full) OA adaption, the larger is the (Hybrid or Full) OA share of the university.

In 2020, and especially for Hybrid OA, a second factor comes into play, namely transformative agreements, understood as part of state regulation here. The share of the publication output of universities covered by such contracts is a factor that adds a considerable amount to the explanatory power of the regression models. Even though transformative agreements are met with scepticism for a number of reasons, including the costs, distributional effects within the German research system and their focus on the three largest publishers that have contributed a lot to the serials crisis in the past, the analysis shows that they are an effective means for the OA transformation with impact on the landscape as a whole.

In contrast, all variables that reflected the infrastructural support for OA on the level of universities and are part of managerial self-governance turned out to be non-significant or did not contribute much to the explained variance. This result should not be interpreted in the sense that infrastructure and support cannot improve the OA share of individual universities. However, the effects are too small to manifest themselves on the level of the entire German university landscape. By contextualising the quantitative analysis with evidence from expert interviews with OA officers from a sample group of German universities, the background about the non-significance of different variables of the managerial self-governance could be explored. In the case of the publication fund, the results suggest that the additional funds are primarily being used by scientists to replace money for APCs from other sources, while for OA policies the interviews show that they have a legitimating function in the first place, instead of directly influencing the OA share. For other variables like 'OA-officer', the provision of OA (rights) information and OA-training and awareness activities, it could not be decided if the variables actually do not influence the journal-based OA share or if they are not significant because of an oversimplified operationalisation in this regression analysis.